\documentclass[3p,twocolumn,preprint,amsmath,amssymb,floatfix]{elsarticle}

\usepackage{graphicx}
\usepackage{epsfig}
\usepackage{dcolumn}
\usepackage{bm}
\usepackage{tabularx}

\newcommand{\bei}{\begin{itemize}}
\newcommand{\eei}{\end{itemize}}
\newcommand{\beq}{\begin{equation}}
\newcommand{\eeq}{\end{equation}}
\newcommand{\beqn}{\begin{eqnarray}}
\newcommand{\eeqn}{\end{eqnarray}}
\newcommand{\beqns}{\begin{eqnarray*}}
\newcommand{\eeqns}{\end{eqnarray*}}

\newcommand{\tabref}[1]{Table~\ref{tab:#1}}

\newcommand{\equaref}[1]{Eq.~(\ref{eq:#1})}

\newcommand{\secref}[1]{Section~\ref{sec:#1}}
\newcommand{\twosecref}[2]{Sections~\ref{sec:#1} and~\ref{sec:#2}}

\newcommand{\figref}[1]{Fig.~\ref{fig:#1}}

\newcommand{\muev} {\mu\rm{eV}}

\begin{document}

\title{Design and performance of the ADMX SQUID-based microwave receiver}

\author[llnl]{S.J. Asztalos\fnref{fn1}, G. Carosi, C. Hagmann, D. Kinion and K. van Bibber\fnref{fn2}}
\fntext[fn1]{Currently at XIA LLC, 31057 Genstar Rd., Hayward CA, 94544.}
\fntext[fn2]{Currently at Naval Postgraduate School, Monterey CA, 93943.}
\address[llnl]{Lawrence Livermore National Laboratory, Livermore, California, 94550}
\author[uw]{ M. Hotz, L. J Rosenberg, G. Rybka, A. Wagner}
\address[uw]{University of Washington, Seattle, Washington 98195}
\author[uf]{J. Hoskins, C. Martin, N.S. Sullivan and D.B. Tanner}
\address[uf]{University of Florida, Gainesville, Florida 32611}
\author[nrao]{R. Bradley}
\address[nrao]{National Radio Astronomy Observatory, Charlottesville, Virginia 22903}
\author[ucb]{John Clarke}
\address[ucb]{University of California and Lawrence Berkeley National Laboratory, Berkeley, California 94720}

\date{\today}

\begin{abstract}
The Axion Dark Matter eXperiment (ADMX) was designed to detect ultra-weakly interacting relic axion particles by searching for their conversion to microwave photons in a resonant cavity positioned in a strong magnetic field. Given the extremely low expected axion-photon conversion power we have designed, built and operated a microwave receiver based on a Superconducting QUantum Interference Device (SQUID). We describe the ADMX receiver in detail as well as the analysis of narrow band microwave signals. We demonstrate the sustained use of a SQUID amplifier operating between 812 and 860 MHz with a noise temperature of 1 K. The receiver has a noise equivalent power of $1.1\times 10^{-24} {\rm W}/{\sqrt{\rm Hz}}$ in the band of operation for an integration time of $1.8\times 10^3$ s.  


\end{abstract}



\maketitle

\section{Introduction}

The Axion Dark Matter eXperiment (ADMX) uses a tunable microwave cavity positioned in a high magnetic field to detect photons from resonantly converted dark matter axions~\cite{Cavity_idea,Cavity_idea_2}. The axion dark matter is extremely weakly coupled to matter and electromagnetic fields so that the power deposited in the cavity from their conversion to photons may be as low as $10^{-24}$ W. The successful detection of this dark matter candidate necessitates a chain of ultra-low noise amplifiers with high gain over the tunable bandwidth of the cavity, between about 300 MHz and 1 GHz. This receiver has already been used to set the most sensitive limits on axion-photon coupling in the $\muev$ mass range~\cite{Squid_results}. 

In addition to axions, many theories predict other ultra-weakly coupled, light particles~\cite{Jaekel_chameleons,Jaekel_photons} that could be detected with the ADMX receiver. The ADMX cavity and receiver have been used to set some of the most stringent constraints on the couplings of such particles~\cite{Chameleons,HS_Photons} in the $\muev$ mass range.

In this paper we describe the design, operation and performance of an ultra low-noise microwave receiver. The receiver has been operated over 812--860 MHz to conduct an axion dark matter search, with the physical temperature of the cavity and first-stage amplifiers of 1.8 K. The system noise temperature, $T_{\rm NS}$, was 3.3 K. We demonstrate that an amplifier based on a dc Superconducting QUantum Interference Device (SQUID) could be successfully matched to a resonant cavity and operated without loss of performance in the presence of the high magnetic field, as required for the axion search. The operating temperature of the cavity was the same as the earlier ADMX generation~\cite{PRD_2001,NIM_paper} so the system noise temperture was dominated by black body noise in the cavity and the reduction in $T_{\rm NS}$ due to the SQUID was quite modest. With the proper choice of components and cooling to 100 mK, the SQUID based receiver is capable of 200 mK system noise temperatures over 300 MHz--3 GHz and beyond. We present an overview of ADMX including the gain and noise requirements the amplifier chain must meet to provide sensitivity to axions and signals from other weakly coupled particles in~\secref{Overview}. A detailed description of the amplifiers follows in~\twosecref{SQUID}{HFET}. In~\secref{Analysis} we describe the search for a generic signal contained within a $0.12$ Hz bandwidth given the raw power spectrum produced by the receiver. Finally, in~\secref{Summary} we summarize our results and discuss the performance of an upgraded receiver.  

\section{Experimental Overview}
\label{sec:Overview}

Here we outline the ADMX receiver chain, summarizing noise requirements and the components of the receiver. The Nyquist resolution of the receiver is 0.12 Hz, providing sensitivity to narrow band signals generated by a variety of physics phenomena. During the nominal axion search, ADMX collects power in each 0.12 Hz bin for a total duration of $1.8\times 10^3$ s so that a signal source producing as little as $1.9\times 10^{-24}$ W in a given bin could be detected with $5~\sigma$ significance with the achieved system noise temperature of 3.3~K. A block diagram of the ADMX receiver is shown in~\figref{Schematic}.

Thermal backgrounds limit the sensitivity of ADMX to power deposited in the cavity. The signal-to-noise ratio ($S/N$) expected from the output of the receiver is given by the radiometer equation~\cite{Dicke_radiometer},

\begin{equation}
\label{eq:Radiometer}
{S\over N} = \frac{P_S}{k_B T_{\rm NS}}\sqrt{\frac{t}{b}}. 
\end{equation}

\noindent Here $P_S$ is the expected signal power in a bandwidth $b$ integrated for a duration $t$ and $k_B$ is the Boltzmann constant. The system noise temperature for a chain of amplifiers is given by~\cite{friis1944}

\begin{equation}
\label{eq:friis}
T_{\rm NS} = T_P + T_{{\rm N}_1} + \sum_{i=2}^{n}{\frac{T_{{\rm N}_i}}{\prod_{j=1}^{i-1}G_j}}, 
\end{equation}

\noindent where $T_P$ denotes the physical temperature of the cavity, $T_{{\rm N}_i}$ is the noise temperature, and $G_i$ is the gain, respectively, of the $i$th amplifier. From~\equaref{friis} we derive the requirement for early stage amplification with high gain and a noise temperature comparable to $T_P$.  

\subsection{Receiver Outline}
\label{sec:RC}

\begin{figure}[htb]
  \begin{center}  
    \epsfig{file=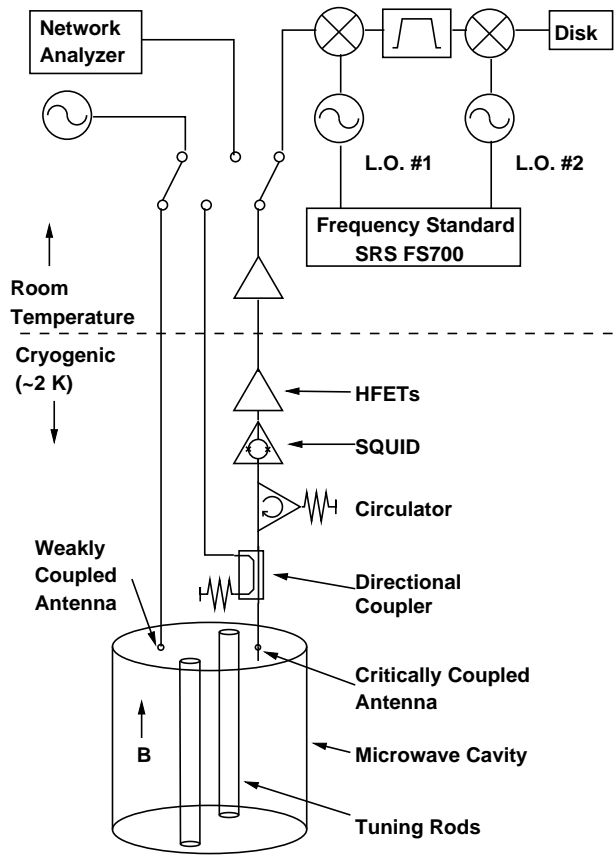, width=80mm}
    \caption{Diagram of the ADMX receiver and electronics. Signals originating in the cavity are amplified by a SQUID and two HFETs maintained at 1.8 K and two post-amplifiers maintained at 300 K before being mixed down to audio frequencies and written to disk. The resonant frequency of the cavity can be adjusted by the position of metal or dielectric rods within.}
    \label{fig:Schematic}
  \end{center}
\end{figure}

The microwave cavity is a copper-plated, stainless-steel cylinder 0.4 m in diameter and 1 m in length~(see~\figref{Schematic}). The unloaded quality factor of the cavity is designed to reach $2\times10^{5}$ at a resonant frequency of 500 MHz. Poor plating of the cavity used in this report limited our quality factor to $2\times10^{4}$, yielding a receiver bandwidth of 25~kHz. The cavity resonant frequency is tuned over a range between 300 and 900 MHz by moving metal or dielectric rods hung within the cavity. The metal rods increase the frequency when moved toward the center of the cavity while dielectric rods decrease the frequency~\cite{Hagmann}. The cavity is maintained at a physical temperature of 1.8 K with pumped liquid helium. An external magnetic field induces axion-photon conversion, requiring the cavity to be placed in a superconducting solenoid supplying a 7.6 T central field~\cite{NIM_paper}. The placement and shielding of the sensitive early stage amplifiers within this high magnetic field are discussed in~\twosecref{SQUID}{HFET}.    

Readout of the cavity is provided by a critically-coupled coaxial antenna inserted into the top plate of the cavity. As shown in~\figref{Schematic}, a second, weakly-coupled antenna is used to inject power into the cavity to determine transmission response. The first stage amplifier is a SQUID load-matched to $50~\Omega$ via a terminated circulator. The SQUID amplifier provides 10 dB of gain with a noise temperature of 1 K when maintained at a physical temperature of 1.8 K. The second stage amplifiers consist of a pair of balanced GaAs Hetero-structure Field-Effect Transistors (HFETs). These amplifiers have noise temperatures of about 4 K and provide a combined gain of 34 dB. After initial amplification by the cryogenic amplifiers the signal is coupled to two room temperature commercial post-amplifier which provide an additional 60 dB of gain. Next, an image-reject mixer (MITEQ IRM045-070-10.7) using a tunable local oscillator (HP 8657A) mixes the signal down to an intermediate frequency (IF) of 10.7~MHz. The signal is then sent through two IF amplifiers each providing 30 dB of gain and a crystal band-pass filter with bandwidth of 30 kHz. Finally, the signal is mixed down to audio frequencies, centered at 20 kHz, before being sent to a 15.5 bit analog-to-digital converter. The Nyquist resolution of the receiver is limited by the phase noise of the frequency standard (SRS FS700) to 0.12 Hz. The gain contributions, noise temperatures, and effective noise contributions of each component are summarized in~\tabref{Receiver}.

\begin{table}[h]
\begin{tabular}{lccc} \hline
Component      & Gain~(dB) & $T_{\rm N}$~(K) & $T_{\rm NS}$~(K) \\
               &           &              &                   \\ 
Cavity         &  -        & 1.8          &  1.8\\ 
SQUID          & 10        & 1.0          &  1.0\\ 
HFET 1         & 17        & 4.0          &  0.4\\ 
HFET 2         & 17        & 4.0          &  0.024\\ 
Post Amplifier & 60        & 100          &  0.004\\ 
Cable          & -6        & 300          &  -\\ 
Image Reject Mixer & -7    & -            &  -\\ 
1st IF Amplifier & 30      & -            &  -\\ 
Crystal Filter & -3        & -            &  -\\ 
2nd IF Amplifier & 30      & -            &  -\\ 
IF - AF Mixer  & -7        & -            &  -\\ \hline
Total          & 141       & -            &  3.3\\ \hline
\end{tabular}
\caption{Noise temperatures and gains contributed by each component of the receiver chain. The ADMX receiver provides 141 dB of gain with a system noise temperature $T_{NS} = 3.3$ K.}
\label{tab:Receiver}
\end{table}

\section{SQUID Amplifiers}
\label{sec:SQUID}

\begin{figure}[h!]
  \begin{center}
    \includegraphics[width=8cm]{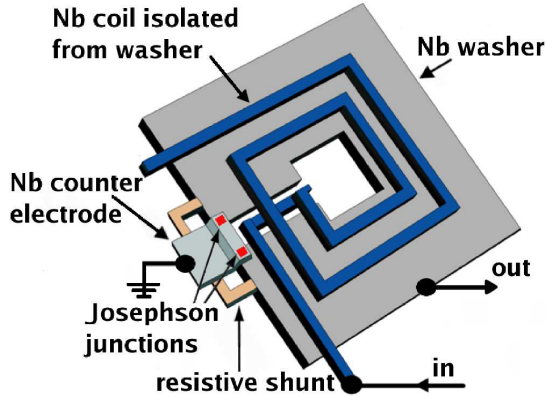}
    \caption{\label{fig:SQUID_diagram} Illustration of a typical MSA and components. Components are not drawn to scale.}
  \end{center}
\end{figure}

\begin{figure}[h!]
  \begin{center}
    \includegraphics[width=8cm]{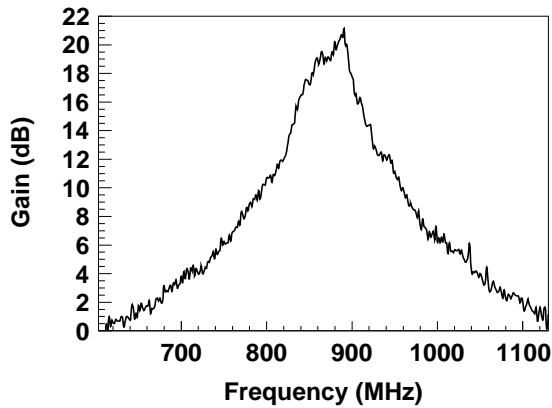}
    \caption{\label{fig:SQUID_gain} Gain (dB) vs. frequency (MHz) for a typical MSA.}
  \end{center}
\end{figure}

The first amplifier in the ADMX receiver chain consists of a dc SQUID~\cite{clarke_squids} of the type described in Ref.~\cite{muck_clarke}. The dc SQUID consist of two Josephson tunnel junctions connected in parallel on a superconducting loop. Each junction is resistively shunted to eliminate hysteresis on the its current-voltage (I-V) characteristic. In ADMX, the SQUID consists of a $1\times 1~{\rm mm}^2$ Nb washer containing a $0.2\times 0.2~{\rm mm}^2$ hole connected to an outer edge by a slit~(\figref{SQUID_diagram}). A Josephson junction is deposited on each side of the slit; the two Nb counter electrodes are connnected together to complete the loop. The dc SQUID is operated with a constant current bias, $I_b$, that produces a voltage of typically 10~$\mu$V. When the magnetic flux $\Phi$ threading the loop is varied, this voltage oscillates with period $\Phi_0=h/2e$; $h$ is Planck's constant and $e$ is the charge of the electron. The SQUID is operated at or near a flux bias of $(n\pm\frac{1}{4})\Phi_0$ for integers $n$, where the transfer coefficient $V_{\Phi}=|\partial V/\partial\Phi|_{I_b}$, is a maximum. Microwave frequency signals are coupled into the SQUID by means of a microstrip consisting of a square Nb coil deposited on the loop with an intervening insulating layer. The signal is connected between the square washer and the outermost turn of the coil~(\figref{SQUID_diagram}). In this way, the coil and SQUID form a resonator permitting high gain at microwave frequencies. We operated the microstrip SQUID amplifier (MSA) at frequencies between 812 and 860 MHz. The gain profile for a typical MSA tuned for operation near 880 MHz is shown in~\figref{SQUID_gain}.

For a tuned input circuit, one can show that the noise temperature of an optimized MSA is given by~\cite{clarke_squids}

\begin{equation}
  \label{eq:MSAnoise}
  T_N\approx\frac{42\nu T}{V_{\Phi}},
\end{equation} 

\noindent where $\nu$ is resonant frequency and $T$ is the physical temperature of the MSA. In the absence of spurious heating effects~\cite{clarke_temp}, $T_N$ decreases linearly with temperature, and may ultimately approach the quantum limit~\cite{muck_clarke}. A noise temperature of $52\pm20$ mK at 538 MHz has been measured in Ref.~\cite{muck_clarke_noise} for an MSA cooled to 20 mK. The quantum limited noise temperature at this frequency is 26 mK~\cite{clarke_note}. 

\begin{figure}[h]
  \begin{center}
    \includegraphics[width=8cm]{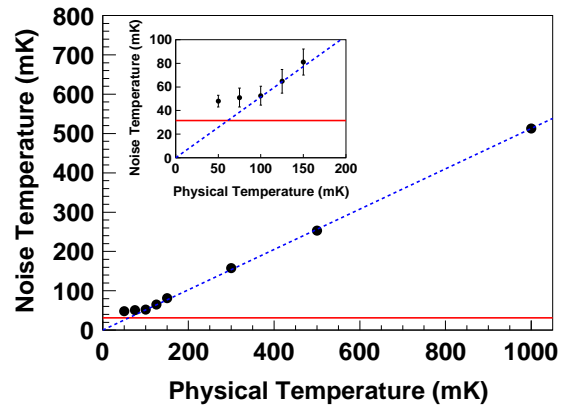}
    \caption{\label{fig:SQUID_noise} Minimum measured noise temperature, $T_N$, vs. physical temperature $T$. The dashed line indicates the proportionality of $T_N$ to $T$ given by~\equaref{MSAnoise} and the solid line represents the quantum limited noise temperature $T_Q = h\nu/k_B$. The inset shows the five lowest physical temperatures.}
  \end{center}
\end{figure}

Measurements of the noise temperature of a representative SQUID used in the ADMX receiver as a function of physical temperature are shown in~\figref{SQUID_noise}. These measurements were performed independently of the ADMX receiver using a separate measurement system.    

The benefit of the low noise temperature of the SQUID can be realized only if its physical temperature is maintained at or near that of the cavity and if transmission losses between the cavity and SQUID are negligible. This requires the SQUID to be physically close to the cavity, within the region of high magnetic field required for axion searches. This placement is necessary despite the extreme sensitivity of the SQUID that requires it to be operated in a ambient magnetic field stable to within 1 nT. We achieve a locally low, stable magnetic field environment for the SQUID with a combination of a field-cancellation magnet, cryoperm shielding, and superconducting shielding. The SQUID is located 0.5 m above the cavity within the field-free region of the cancellation magnet and is cooled via four 1 m long copper braids attached to the cavity.

\begin{figure}[h]
  \begin{center}
    \includegraphics[width=7.5cm]{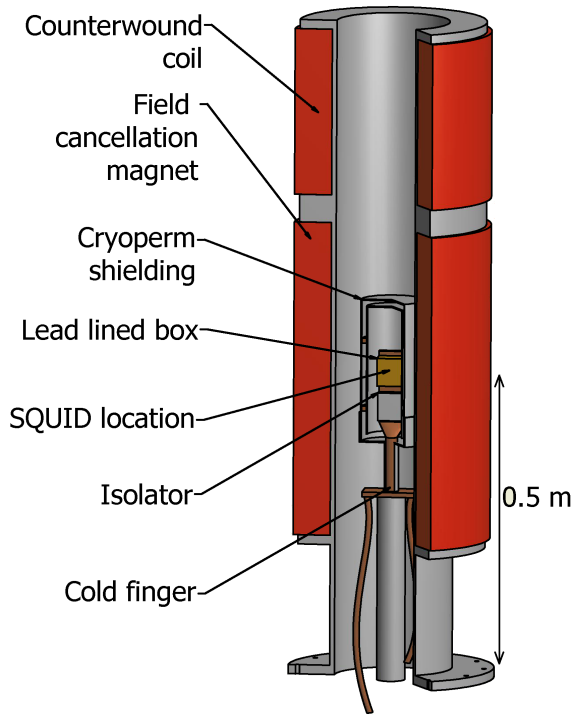}
    \caption{\label{fig:shielddiagram} Illustration of the SQUID positioned within the field-cancellation magnet. The SQUID is shielded within two cryoperm shields and a lead-lined box. The box is cooled via copper braids attached to the cavity.}
  \end{center}
\end{figure}

\begin{figure}[h]
  \begin{center}
    \includegraphics[angle=270,width=8cm]{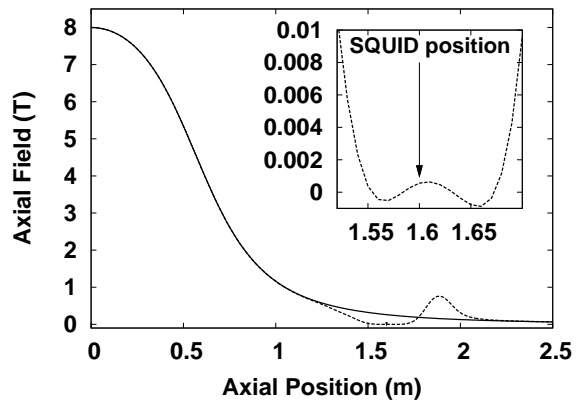}
    \caption{\label{fig:buckingcoilfield} Calculated axial magnetic field with (dashed line) and without (solid line) bucking coil, showing the SQUID in the 1 mT region.}
  \end{center}
\end{figure}

The field-cancellation magnet is a 0.914 m long superconducting coil with a 0.152 m diameter bore, producing a field-free region 0.13 m in length centered 0.5 m above the cavity. The coil of the magnet contains a counter wound section to minimize mutual inductance with the main magnet. This design permits the field-cancellation magnet to be energized independently of the main magnet and produces less than 400 N net force between the two magnets. The field-cancellation magnet reduces the field in the region near its axial center to a calculated value less than 1 mT as shown in~\figref{buckingcoilfield}. Inside the field-cancellation magnet, two concentric cylinders of cryoperm shielding reduce the local field to about $0.1~\mu$T. The SQUID itself is housed inside a lead-lined box at the center of the cryoperm cylinders. The superconducting lead plating stabilizes the magnetic field to within $0.5$~nT or $0.05~\Phi_0$, when cooled below its transition temperature. The flux bias of the SQUID required adjustment when large deviations of the ambient field occured. The placement of the SQUID within the field-cancellation magnet and shielding is shown in~\figref{shielddiagram}.

\begin{figure}[h]
  \begin{center}
    \includegraphics[angle=270,width=8cm]{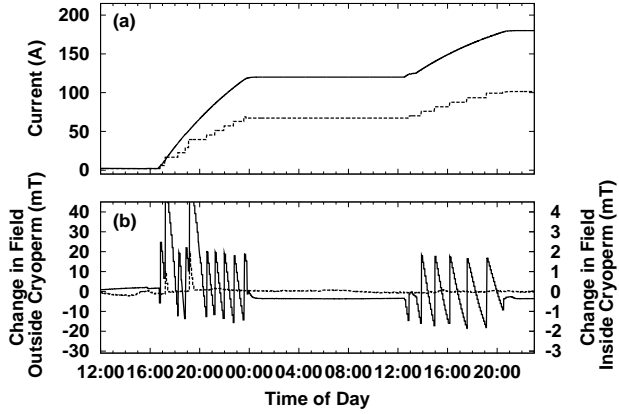}
    \caption{\label{fig:ramp} Typical magnet ramp-up procedure. (a): Main magnet current (solid) and bucking coil current (dashed).  (b): Field outside cryoperm shield (solid) and field inside cryoperm shield (dashed).}
  \end{center}
\end{figure}

While the main magnet is brought to its operating field of 7~T, the current in the field cancellation magnet is adjusted to keep the field below 25~mT outside the passive shielding for the SQUID and isolator. The field measured by Hall sensors outside the superconducting lead shield is used to control the cancellation magnet current in steps to provide continuous field cancellation when energizing the magnet, shown in~\figref{ramp}.

\section{HFET Amplifiers}
\label{sec:HFET}

Following the initial SQUID preamplifier, two stages of cryogenic amplification are provided by balanced GaAs HFETs. A cryogenic RF coax cable brings the signal from the SQUID amplifier in the field free region to two cooled HFET amplifiers and then to room-temperature RF amplifiers located just outside the ADMX vacuum system. The HFET amplifiers were developed at the National Radio Astronomy Observatory (NRAO); details of their design can be found in~\cite{Bradley_1}.

\begin{figure}[h]
  \hfill
    \begin{center}  
      \epsfig{file=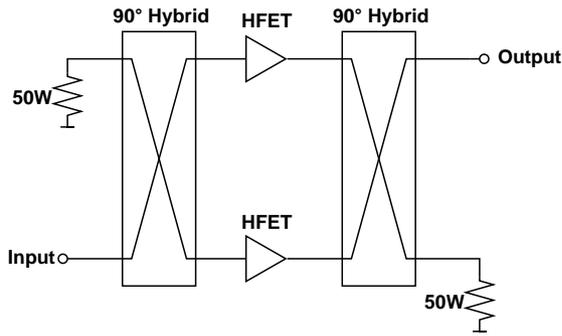, width=75mm}
      \caption[Schematic of balanced HFET amplifier.]{Illustration of a balanced HFET amplifier.}
      \label{fig:Balanced}
    \end{center}
\end{figure}

Each amplifier consists of two matched single-ended HFETs along with two four-port hybrid couplers arranged in the balanced configuration shown in~\figref{Balanced}. The input signal is split at the first hybrid and one branch is given a 90$^{\circ}$ phase shift. Both signals are then amplified and passed through a second hybrid where the output signals of both HFETs are brought back into phase. The key feature of this design is that reflections at the input stage of both HFETs are 180$^{\circ}$ out of phase, thus allowing the balanced amplifier to match the input impedance over a broad frequency range.

\begin{figure}[h]
  \begin{center}  
    \includegraphics[angle=270,width=8cm]{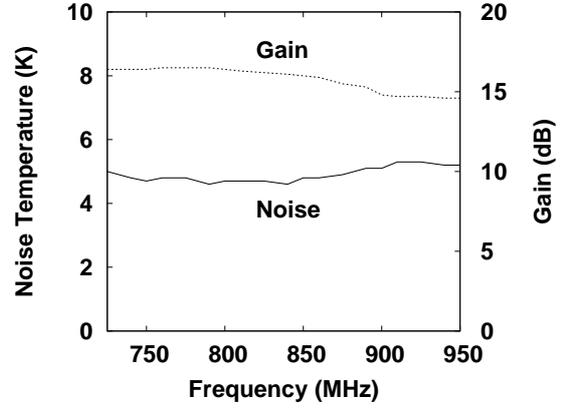}
    \caption[HFET gain (dashed) and noise (solid) vs. frequency.]{HFET gain (dashed) and noise (solid) vs. frequency.}
    \label{fig:HFET_gain_noise}
  \end{center}
\end{figure}

The first and second stage HFET amplifiers had gains of 17--18 dB at 4.2 K with peaks at 800 MHz and 650 MHz, respectively. In the frequency range of interest (800--900 MHz) the combined gain was approximately 30 dB. The contributions of the HFETs to the noise temperature [~\equaref{friis} and~\tabref{Receiver}] were 400 mK and 24 mK respectively. The gain and noise temperature for a typical HFET amplifier operated at 4.2 K are shown in~\figref{HFET_gain_noise} as a function of frequency. 

Initial HFET amplifier testing showed a sensitivity to magnetic fields believed to be due to the Lorentz force affecting the HFET transconductance. This effect is highly dependent on the orientation of the amplifiers relative to the field~\cite{Bradley_2}. As a result, care was taken to orient both HFET amplifiers vertically so that the direction of electron propagation in the channels were parallel to the magnetic field. The HFETs were attached to a copper plate anchored to one of the three cavity support brackets. The HFETs dissipated roughly 10 mW, and were thermally isolated from the bracket with nylon spacers to prevent local heating of the cavity. A resistive temperature sensor attached to the copper plate allowed for accurate physical temperature measurements. 



\section{Data Analysis}
\label{sec:Analysis}

\begin{figure}[h!]
  \begin{center}
    \includegraphics[width=8cm]{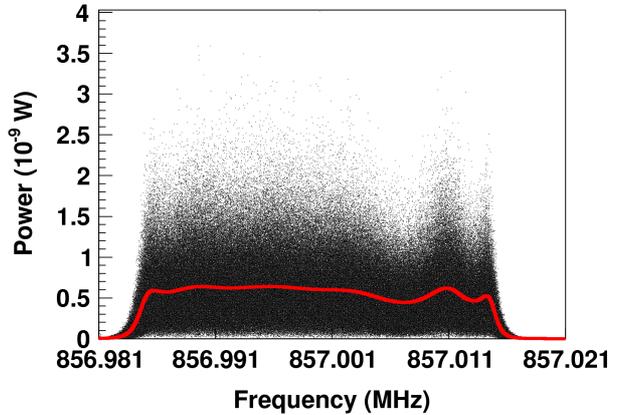}
    \caption{\label{fig:rawspectrum} Power spectrum from single 71.5 second integration overlaid with the smoothed daily average. The primary structure is from the transfer function of the crystal filter.}
  \end{center}
\end{figure}

As previously stated, the axion signal is a power excess within a less than 1 kHz bandwidth, stationary in frequency over the timescale of minutes. The response of the amplifiers and filters in the receiver produces broadband structure over the 25 kHz response, shown in~\figref{rawspectrum}, that we remove with an average power spectra shape computed every 24 hrs. Data are sampled at 80~kHz for 71.5~s at 15.5 bits resolution and Fourier transformed into a power spectrum. After each 71.5~s integration the tuning rods are adjusted to change the resonant frequency of the cavity by 1 kHz so that each 0.12 Hz frequency bin is integrated for a total of $1.8\times 10^3$~s. 

\begin{figure}[h!]
  \begin{center}
    \includegraphics[width=8cm]{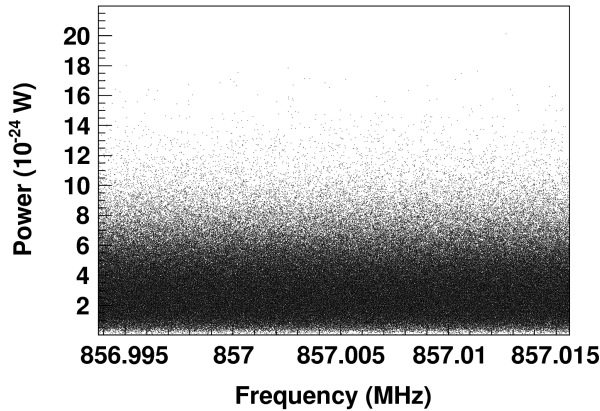}
    \caption{\label{fig:power} Power spectrum data after removal of the receiver transfer function, scaled to expected thermal power.}
  \end{center}
\end{figure}

The receiver transfer function is estimated each day by averaging and smoothing all of the power spectra from a day. Each power spectrum in a given day is normalized by the transfer function to remove the effects of variable amplifier and filter response, and scaled by the expected thermal power. The observed power spectrum for a typical 71.5~s integration is shown in~\figref{power}.

\noindent We average subsequent spectra and find that the power sensitivity $\delta P/P$ improves as $1/\sqrt{t}$ for $t$ up to $10^6$~s.

\begin{figure}[h!]
  \begin{center}
    \includegraphics[width=8cm]{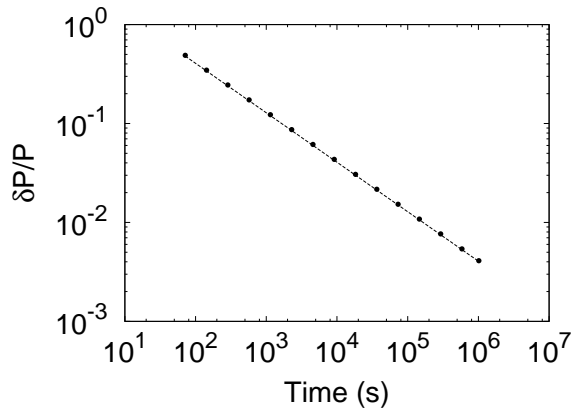}
    \caption{\label{fig:noise_averaging} Power sensitivity $\delta P/P$ as a function of integration time up to $10^6$~s. Data are shown as dots. The dashed line has a slope of $-1/2$.}
  \end{center}
\end{figure}

\section{Summary and Outlook}
\label{sec:Summary}

We operated the ADMX with a SQUID-based microwave receiver in the frequency range 812-860 MHz. The receiver has a noise equivalent power of $1.1\times 10^{-24} {\rm W}/{\sqrt{\rm Hz}}$ in the band of operation for an integration time of $1.8\times 10^3$ s. We maintain a low, stable magnetic field environment for the SQUID with the combination of a field cancellation magnet and cryoperm shielding. The operation of the SQUID MSA in the presence of a high magnetic field is broadly applicable to searches for axions and other exotic particles~\cite{Chameleons,HS_Photons}, dramatically improving the scan rate and coupling sensitivity. 

A planned upgrade of ADMX will employ a dilution refrigerator to cool the cavity and SQUID to 100 mK resulting in a more than tenfold improvement in both the physical operating temperature and noise temperature. The use of moderately tunable SQUID MSAs~\cite{muck_clarke_tune} with higher operating frequencies~\cite{muck_welzel} will permit searches for weakly coupled physics over an order of magnitude in microwave energies. This upgrade will enable ADMX to search for a large variety of physics with an unprecedented sensitivity. 

\section{Acknowledgments}

The ADMX collaboration gratefully acknowledges support by the U.S. Department of Energy, Office of High Energy Physics under contract numbers DE-FG02-96ER40956 (University of Washington), DE-AC52-07NA27344 (Lawrence Livermore National Laboratory), and DE-FG02-97ER41029 (University of Florida). Additional support was provided by Lawrence Livermore National Laboratory under the LDRD program. Development of the SQUID amplifier (JC) was supported by the Director, Office of Science, Office of Basic Energy Sciences, Materials Sciences and Engineering Division, of the U.S. Department of Energy under Contract No. DE-AC02-05CH11231.

\bibliographystyle{h-physrev}
\bibliography{admx_receiver}

\begin{thebibliography}{10}

\bibitem{Cavity_idea}
P.~Sikivie,
\newblock Phys. Rev. Lett. {\bf 51}, 1415 (1983).

\bibitem{Cavity_idea_2}
P.~Sikivie,
\newblock Phys. Rev. D {\bf 32}, 2988 (1985).

\bibitem{Squid_results}
S.~Asztalos {\em et~al.},
\newblock Phys. Rev. Lett. {\bf 104}, 041301 (2010).

\bibitem{Jaekel_chameleons}
J.~Jaeckel, E.~Masso, J.~Redondo, A.~Ringwald, and F.~Takahashi,
\newblock Phys. Rev. D {\bf 75}, 013004 (2007).

\bibitem{Jaekel_photons}
J.~Jaeckel, E.~Masso, and A.~Ringwald,
\newblock Phys. Lett. B {\bf 659}, 509 (2008).

\bibitem{Chameleons}
G.~Rybka {\em et~al.},
\newblock Phys. Rev. Lett. {\bf 105}, 051801 (2010).

\bibitem{HS_Photons}
A.~Wagner {\em et~al.},
\newblock Phys. Rev. Lett. {\bf 105}, 171801 (2010).

\bibitem{PRD_2001}
S.~Asztalos {\em et~al.},
\newblock Phys. Rev. D {\bf 64}, 092003 (2001).

\bibitem{NIM_paper}
H.~Peng {\em et~al.},
\newblock Nucl. Instrum. Methods Phys Res., Sect. A {\bf 444}, 569 (2000).

\bibitem{Dicke_radiometer}
R.~Dicke,
\newblock Rev. Sci. Instrum. {\bf 17}, 268 (1946).

\bibitem{friis1944}
H.~T. Friis,
\newblock Proc. of the IRE {\bf 32}, 419  (1944).

\bibitem{Hagmann}
C.~Hagmann, P.~Sikivie, N.~Sullivan, D.~Tanner, and S.-I. Cho,
\newblock Rev. Sci. Instrum. {\bf 61}, 1076 (1990).

\bibitem{clarke_squids}
J.~Clarke, A.~Lee, M.~M\"{u}ck, and P.~Richards,
\newblock {SQUID} voltmeters and amplifiers,
\newblock in {\em The SQUID Handbook Vol. II: Applications of SQUIDs and SQUID
  systems}, pp. 1--93, 2006.

\bibitem{muck_clarke}
M.~M\"{u}ck, M.~Andr\'{e}, J.~Clarke, J.~Gail, and C.~Heiden,
\newblock App. Phys. Lett. {\bf 72}, 2885 (1998).

\bibitem{clarke_temp}
F.~Wellstood, C.~Urbina, and J.~Clarke,
\newblock Phys. Rev. B {\bf 49}, 5942 (1994).

\bibitem{muck_clarke_noise}
M.~M\"{u}ck, J.~Kycia, and J.~Clarke,
\newblock App. Phys. Lett. {\bf 78}, 967 (2001).

\bibitem{clarke_note}
We note that a SQUID amplifier can actually achieve the quantum limit only when
  operated slightly below resonance~\cite{clarke_squids}.

\bibitem{Bradley_1}
R.~Bradley {\em et~al.},
\newblock Nuclear Physics B (Proc. Suppl.) {\bf 72}, 137 (1999).

\bibitem{Bradley_2}
E.~Daw and R.~Bradley,
\newblock J. Applied Physics {\bf 82}, 1925 (1997).

\bibitem{muck_clarke_tune}
M.~M\"{u}ck, M.~Andr\'{e}, J.~Clarke, J.~Gail, and C.~Heiden,
\newblock App. Phys. Lett. {\bf 75}, 3545 (1999).

\bibitem{muck_welzel}
M.~Muck, C.~Welzel, and J.~Clarke,
\newblock Appl. Phys. Lett. {\bf 82}, 3266 (2003).

\end{thebibliography}

\end{document}